\documentclass[sigconf,svgnames]{acmart}

\usepackage{booktabs} 
\usepackage{threeparttable}
\usepackage{colortbl} 
\usepackage{color}
\usepackage{textcomp}
\usepackage{soul} 
\usepackage{amsmath}
\usepackage{comment}
\usepackage{multirow}
\usepackage{algorithm} 
\usepackage{algpseudocode} 

\newtheorem{myDef}{Definition}

\copyrightyear{2018} 
\acmYear{2018} 
\setcopyright{acmcopyright}
\acmConference[CIKM '18]{The 27th ACM International Conference on Information and Knowledge Management}{October 22--26, 2018}{Torino, Italy}
\acmBooktitle{The 27th ACM International Conference on Information and Knowledge Management (CIKM '18), October 22--26, 2018, Torino, Italy}
\acmPrice{15.00}
\acmDOI{10.1145/3269206.3271694}
\acmISBN{978-1-4503-6014-2/18/10}

\begin{document}
\fancyhead{}
\title{Mathematics Content Understanding for Cyberlearning via Formula Evolution Map}

\author{Zhuoren Jiang}
\affiliation{
\institution{$^1$Sun Yat-sen University}
\city{Guangzhou} 
\country{China}}
\affiliation{\institution{$^2$Peking University, Beijing, China}}
\email{jiangzhr3@mail.sysu.edu.cn}

\author{Liangcai Gao}
\authornote{Corresponding author}
\affiliation{
\institution{Peking University}
\city{Beijing} 
\country{China}}
\email{glc@pku.edu.cn}

\author{Ke Yuan}
\affiliation{
\institution{Peking University}
\city{Beijing} 
\country{China}}
\email{yuanke@pku.edu.cn}

\author{Zheng Gao}
\affiliation{
\institution{Indiana University Bloomington}
\city{Bloomington} 
\state{IN}
\country{USA}}
\email{gao27@indiana.edu}

\author{Zhi Tang}
\affiliation{
\institution{Peking University}
\city{Beijing} 
\country{China}}
\email{tangzhi@pku.edu.cn}

\author{Xiaozhong Liu}
\authornotemark[1]
\affiliation{
\institution{$^1$Alibaba Group}
\city{Seattle \& Hangzhou} 
\country{China}}
\affiliation{
\institution{$^2$Indiana University Bloomington}
\city{Bloomington} 
\state{IN}
\country{USA}}
\email{liu237@indiana.edu}

\begin{abstract}
Although the scientific digital library is growing at a rapid pace, scholars/students often find reading Science, Technology, Engineering, and Mathematics (STEM) literature daunting, especially for the math-content/formula. In this paper, we propose a novel problem, ``mathematics content understanding'', for cyberlearning and cyberreading. To address this problem, we create a Formula Evolution Map (FEM) offline and implement a novel online learning/reading environment, PDF Reader with Math-Assistant (PRMA), which incorporates innovative math-scaffolding methods. The proposed algorithm/system can auto-characterize student emerging math-information need while reading a paper and enable students to readily explore the formula evolution trajectory in FEM. Based on a math-information need, PRMA utilizes innovative joint embedding, formula evolution mining, and  heterogeneous graph mining algorithms to recommend high quality Open Educational Resources (OERs), e.g., video, Wikipedia page, or slides, to help students better understand the math-content in the paper. Evaluation and exit surveys show that the PRMA system and the proposed formula understanding algorithm can effectively assist master and PhD students better understand the complex math-content in the class readings. 
\end{abstract}

\settopmatter{printccs=false}


\maketitle

\vspace{-2ex}\section{Introduction}


Over the past decade, while the volume of Science, Technology, Engineering, and Mathematics (STEM) publications has increased dramatically in university and digital libraries, few efforts have been made to help readers, especially junior scholars and graduate students, understand them. From a learning and reading viewpoint, understanding the content (especially the math-content) of scientific publications in STEM remains daunting~\cite{liu2014interactive}. In a survey conducted involving computer science program (35 Master and Ph.D. students), participants rated readings (textbook/publications) in data mining to be difficult (45.71\%) or very difficult (14.29\%). Furthermore, students claimed that math content in the readings was too difficult and inscrutable to understand because of the readers' limited knowledge in statistics and mathematics, i.e., participants believed the mathematical content in the readings to be difficult (51.43\%) or very difficult (22.86\%). Meanwhile, all the participates believe these papers are important, and they hope they could get additional help to better understand the math content in these papers. This survey motivated our thinking about this new problem - \textbf{Mathematics Content Understanding (MCU)}, a.k.a. how can we propose a useful method to assist readers to better understand the math-content in an academic publication. Junior students who struggle with math problem or scholars who want to conduct interdisciplinary research can especially get benefit from this study. To the best of our knowledge, this is the first study investigates the MCU problem. However, MCU faces the following challenges:

\textbf{Math content representation.} There are significant differences between math content (formula in most cases) and natural language. First, the mathematical symbols of a formula are ambiguous. For instance, the variable ``$\alpha$'' or ``$x$'' could represent the same meaning if defined by different scholars. Second, formulae can carry recursive structures while natural language is usually linear in structure. Third, formulae are highly structured and usually presented in a layout form, e.g., \LaTeX{} or \textit{MathML}. Existing text mining method can be hardly used to address MCU problem.

\textbf{Information need shifting.} Student's math information need could potentially shift when facing a complicated formula. For instance, to understand the formula of \textit{``Latent Dirichlet Allocation (LDA)''}, one may need to understand the formulae of \textit{``Dirichlet/Beta distribution''} (component) or even \textit{``Conditional Probability''} (foundation). From the evolutionary viewpoint, these formulae can be considered as the ``ancestors'' of the original formula, and have important auxiliary effects for MCU. However, such information can not be fully extracted from the formula context or citations. 

\textbf{Information access $\neq$ information understanding.} Though traditional formula retrieval models could help user to access the math information. But, understanding information is fundamentally different from accessing information \cite{liu2015scientific}. User need more supportive information to understand the math-content in the publications. For instance, \cite{liu2013answering} showed that cyberlearning resources (i.e., slides or video) can be more helpful (than scholarly publications) for scientific understanding. 

\begin{figure}[htb]\centering \vspace{-2ex}
	\includegraphics[width=0.9\columnwidth]{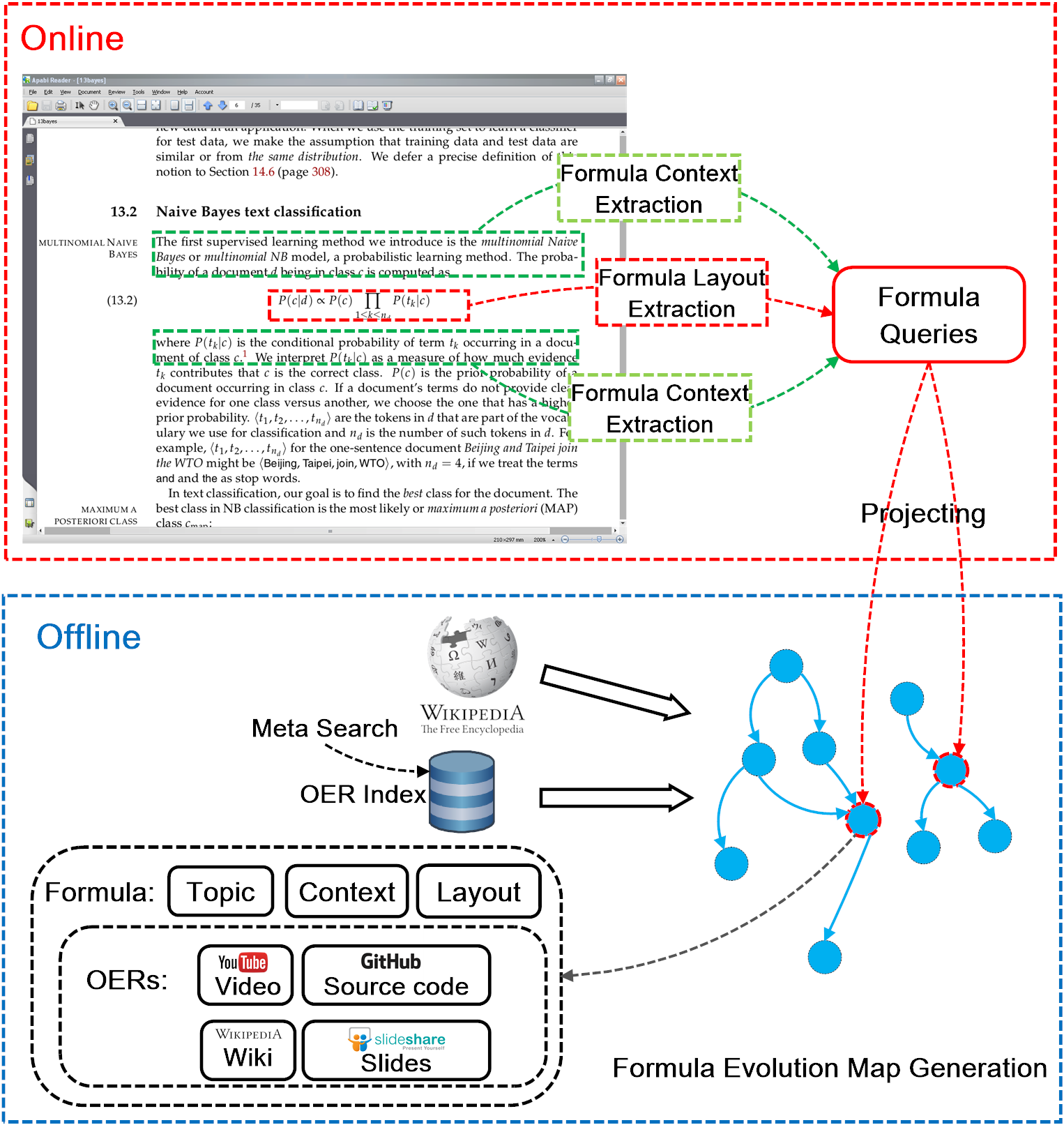} 
	\vspace{-2ex}
	\caption{The whole framework for this study }
	\label{fig:frame}
 \end{figure}

In order to address these challenges, this paper proposes a novel solution, \textbf{MCU via Formula Evolution Map}, in a broad area of information retrieval and education. Although STEM publications generally do not place a premium on writing for readability, in this study, we hypothesize that the formula evolution information can be important to assist readers to better understand the math-content in a paper. As Figure \ref{fig:frame} shows, we investigate the following two processes:

$\bullet$ In an \textit{offline process}: By mining a large number of scientific knowledge-base documents and the associated formulae, from heterogeneous graph and joint embedding perspectives, we generate the \textbf{Formula Evolution Map (FEM)}, which encapsulates the mathematical evolutionary information over time.

$\bullet$ In an \textit{online process}: By leveraging formula layout and context information extraction, we will \textbf{project the user information need (while reading the math-content in a publication) to the FEM}, as well as recommend useful resources to help readers better understand and consume the target publication and associated math content.

For FEM generation, the proposed algorithm explores more than 4 million documents and the associated math/formula/algorithmic information to characterize comprehensive and fundamental formula evolutionary relations in STEM (there are a total of 21,292,157 evolutionary relations on the FEM). To help readers/students better consume the target paper, in the FEM graph index, each formula vertex also associates the formula layout information, target topic (the formula belongs to), and a number of Open Educational Resources (OERs), i.e., video lectures, presentation slides, source codes, and Wikipedia pages, that may help readers to better consume the mathematical content in the online environment. 

Meanwhile, in order to verify the proposed algorithm and cyberlearning hypothesis, we design a novel reading environment (PDF Reader with Math-Assistant, PRMA). When using PRMA system, a reader can easily highlight a formula with the mouse, and the system can automatically project the target formula in the paper to the vertexes on the backend FEM as well as recommend OERs for formula understanding. Evaluation results show that the proposed method is important to help students understand the math-content in a paper, and its potential in cyberlearning is promising. 

\textbf{The contribution of this paper is fourfold.} First, we propose an innovative MCU problem in an education context to help students and junior scholars better consume the STEM publications. Second, a new reading environment is employed to capture student information need while enabling them to highlight the confusing formula of the reading. Third, novel algorithms are proposed to characterize \textit{formula evolution information} in a map by mining a massive scientific knowledge base. Last but not least, an extensive experiment (with 52 participants) is employed to qualitatively and quantitatively validate the proposed formula understanding hypothesis as well as the usefulness of the system and to evaluate the FEM generation quality. As MCU is a newly proposed but important problem, we share the algorithm generated FEM plus massive formula and math-topic information to motivate further investigation.

  \vspace{-2ex}

Based on the experiment participants' feedback, 72.73\% of participants believed the proposed method can provide very useful information for math-understanding and 75.75\% of participants believed the system recommended OERs (especially for videos and slides), comparing with the text content, are much more helpful for math-understanding. Algorithm evaluation also shows that FEM and the associated formula evolution relations are very important for math-understanding (can enhance precision and NDCG significantly).


\vspace{-1ex}\section{Problem Formulation} \label{sec:problem}

As aforementioned, simply return academic papers may not be enough to help junior scholars \cite{liu2013answering}. In this study, we hypothesize that formula evolution information along with OERs can be important to address MCU problem.

\begin{myDef}
\textbf{Formula Evolution Map (FEM)}. FEM is defined as a weighted directed graph $G=(F,R,\tau)$, where $F$ denotes the formula vertex set, and $R\subseteq  F\times F$ denotes the directed evolution relation set. $\omega$ is the relation weight set, denotes formula evolution probabilities. 
\end{myDef}

FEM encapsulates the fundamental and enlightened formula evolution information, which could be especially useful for exploring the development of a formula as well as the details of its components. For instance, formula of ``\textit{Bayes' theorem}'' is the foundation of ``\textit{Naive Bayes classifier}'' formula, and formulae of ``\textit{Gaussian naive Bayes}'' and ``\textit{Multinomial naive Bayes}'' are both the specific forms of general ``\textit{Naive Bayes}''. There are clear evolutionary paths among these formulae. 

\begin{myDef}
\textbf{Mathematics Content Understanding via OER}. From OER recommendation viewpoint, the MCU problem can be defined as a conditional probability \textit{P(OER|info-need)}, i.e., the probability of an OER given a particular math information need, which can be formalized as:
\begin{itemize}
\item \textbf{Input}: A mathematical content (a formula with its context).
\item \textbf{Output}: A list of ranked OERs that could be potentially useful for understanding the target math-content. 
\end{itemize}
\end{myDef}

\begin{myDef}
\textbf{Mathematics Content Understanding via FEM + OER}. Based on prior definitions, we can further integrate FEM into consideration, i.e., \textit{P(OER|info-need) = P(OER|formula) $\ast$ P(formula|info-need)}, where user's math information need can be projected to a formula with its ancestors (vertexes) on FEM for MCU. Then, the projected formulae (on FEM) and their related OERs can help to address MCU problem.
\end{myDef}
\vspace{-2ex}\section{Methodology}\label{sec:method}

In this section, we discuss the research method in detail including: generating the Formula Evolution Map (FEM) offline~(3.1), designing the novel cyberlearning environment to characterize readers' information need when consuming mathematics content in a paper~(3.2), and designing OER recommendation for formula understanding~(3.3).

\vspace{-2ex}\subsection{Formula Evolution Map Generation}

\vspace{-2ex}\begin{figure}[!htb]\centering
	\includegraphics[width=1.0\columnwidth]{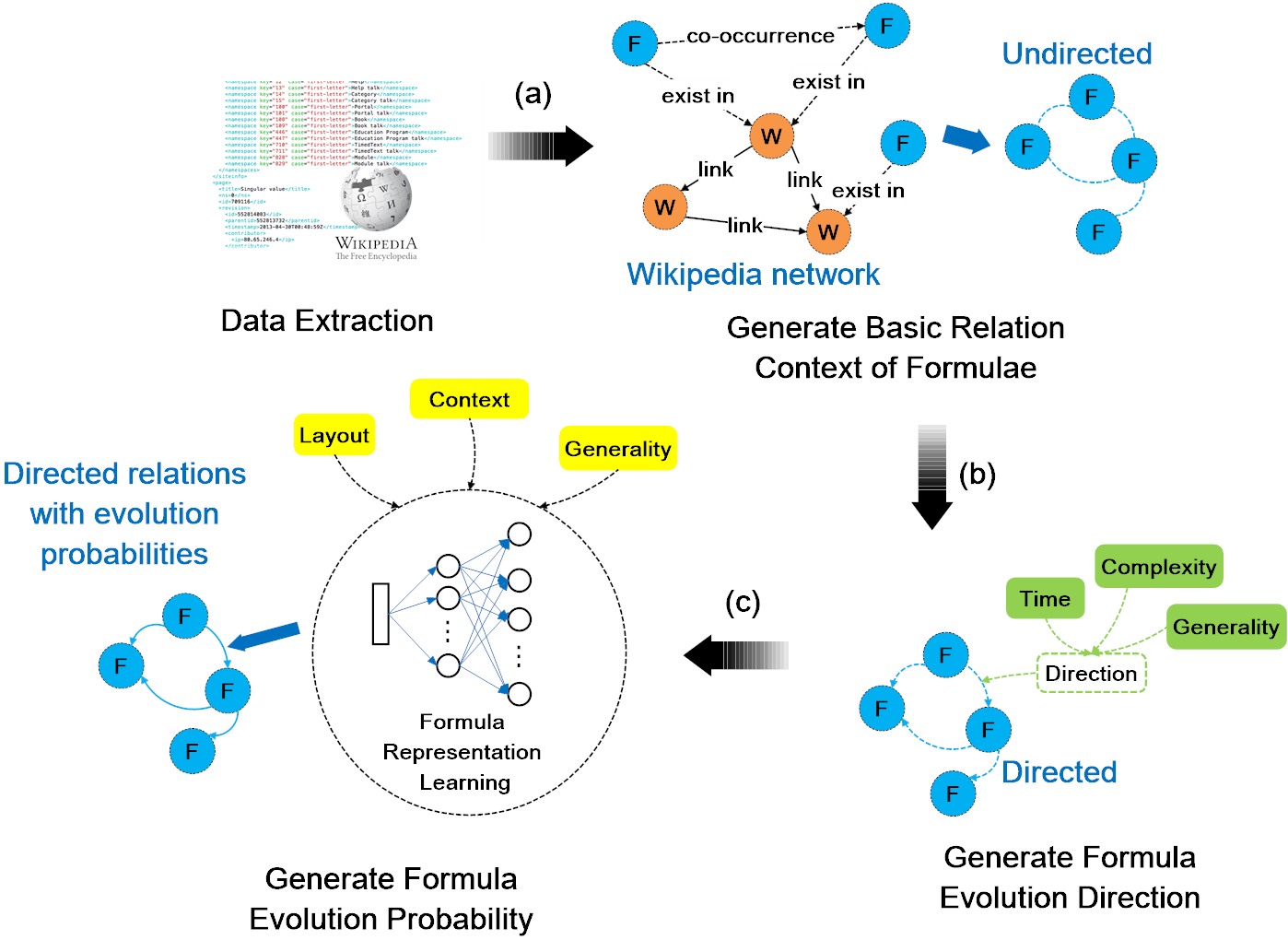} 
	\caption{Formula evolution map generation}\vspace{-2ex}
	\label{fig:femgeneration}
 \end{figure}
 
We generate a Formula Evolution Map (FEM), offline, to interconnect important/fundamental formulae in scientific publications. Note that the formulae in the FEM cannot cover all the formulae in the readings. Instead, it provides the potential to 1) project any formula in any paper to the vertex(es) in FEM, and 2) trace the formula evolution information on FEM for math-understanding. 

To achieve this goal, we employ a large knowledge base, Wikipedia, to generate FEM. There are two reasons we use Wikipedia: first, Wikipedia contains a wealth of mathematical information, including tens of thousands of fundamental formulae (with math-topic and formula layout information); second, Wikipedia provides links between pages, which can be important to generate formula evolution information. In this paper, we use the text information, formula layout information, and link topology information extracted from Wikipedia dump to generate FEM that enables an algorithm to estimate formulae evolution over time. More importantly, the proposed FEM can minimize the noisy formulae information.

The FEM generation process is illustrated in Figure \ref{fig:femgeneration}, which involves three steps: (a) formula evolution relation generation, (b) formula evolution direction determination, (c) formula evolution probability calculation. 

\textbf{(a) Formula Evolution Relation Generation}: By using the hyperlinks between Wikipedia pages and the co-occurrence relationships (in the same Wikipedia page) of formulae, we generate the basic relation context of formulae. The formula evolution relation generation can be modeled as:
\begin{equation}
R(f_{a},f_{b}|W) = \textup{Sgn}_{r} \left ( w_{a},w_{b} \right ) =\left\{\begin{matrix}
1 & Co(w_{a},w_{b})=1 || w_{a}=w_{b}\\ 
0 & otherwise
\end{matrix}\right.
\end{equation}

Here, $R$ is a signum function $\textup{Sgn}_{r}$, that indicates the evolution relation existence between formula $f_{a}$ and formula $f_{b}$ based on the Wikipedia page network $W$. $w_{a}$ is the home page of $f_{a}$, and $w_{b}$ is the home page of $f_{b}$ (extracted from the Wikipedia dump). $Co(w_{a},w_{b}) = 1 $ indicates that there is a hyperlink between $w_{a}$ and $w_{b}$, and $w_{a}=w_{b}$ means $f_{a}$ and $f_{b}$ are hosted in the same page. When $R(f_{a},f_{b}) = 1$, there could be a candidate evolution relation between $f_{a}$ and $f_{b}$.

In the generated undirected formula relation network, each formula is characterized as a vertex with multiple attributes: (1) Wikipedia page title, (2) formula context information (250 characters), and (3) formula layout information.

\textbf{(b) Formula Evolution Direction Determination}: In this study, the evolution direction between formulae are determined by three indicators (assumptions): (1) $\lambda _{t}\left ( f \right )$ , formula birth time (formulae could evolve from past to present); (2) $\lambda _{g}\left ( f \right )$, formula generality (formulae could evolve from fundamental to contextualized); (3) $\lambda _{c}\left ( f \right )$ formula layout complexity (formulae could evolve from simple to complex).

For a formula pair $\left \{f_{a}, f_{b}\right \}$, $R(f_{a},f_{b}|W) = 1$, the evolution direction is first decided by $\lambda _{t}$. However, not every formula has $\lambda _{t}$ attribute. If the ``\textit{birth time}'' of a formula is missing, we use $\lambda _{c}$ to determine the direction. To avoid the uncertainty caused by the layout comparison, if the ratio of the complexity difference $< 0.1$, we use $\lambda _{g}$ as the final direction indicator.

To generate $\lambda _{t}\left ( f \right )$, we use the title of home Wikipedia page to represent the formula, then we use the greedy match algorithm in a large academic paper corpus to find the earliest appearance of the formula. Note that one formula may exist in multiple Wikipedia pages. We use the first appearance time among the Wikipedia concepts as this formula's creation time. The smaller $\lambda _{t}\left ( f \right )$ is, the earlier the formula $f$ appears.

Meanwhile, PageRank \cite{page1999pagerank} is utilized to calculate the formula generality $\lambda _{g}\left ( f \right )$,  for measuring the fundamental level of a formula. The underlying assumption is similar as PageRank: more fundamental formulae are likely to receive more links from other formulae. $\lambda _{g}$ is calculated via a formula-formula graph (generated from the page-page wiki-graph), and each vertex in the graph is a formula. The generality of a formula is voted by the links among formulae. We hypothesize that formulae could evolve from past to the present.

\begin{figure}[!htb]
	\includegraphics[width=0.75\columnwidth]{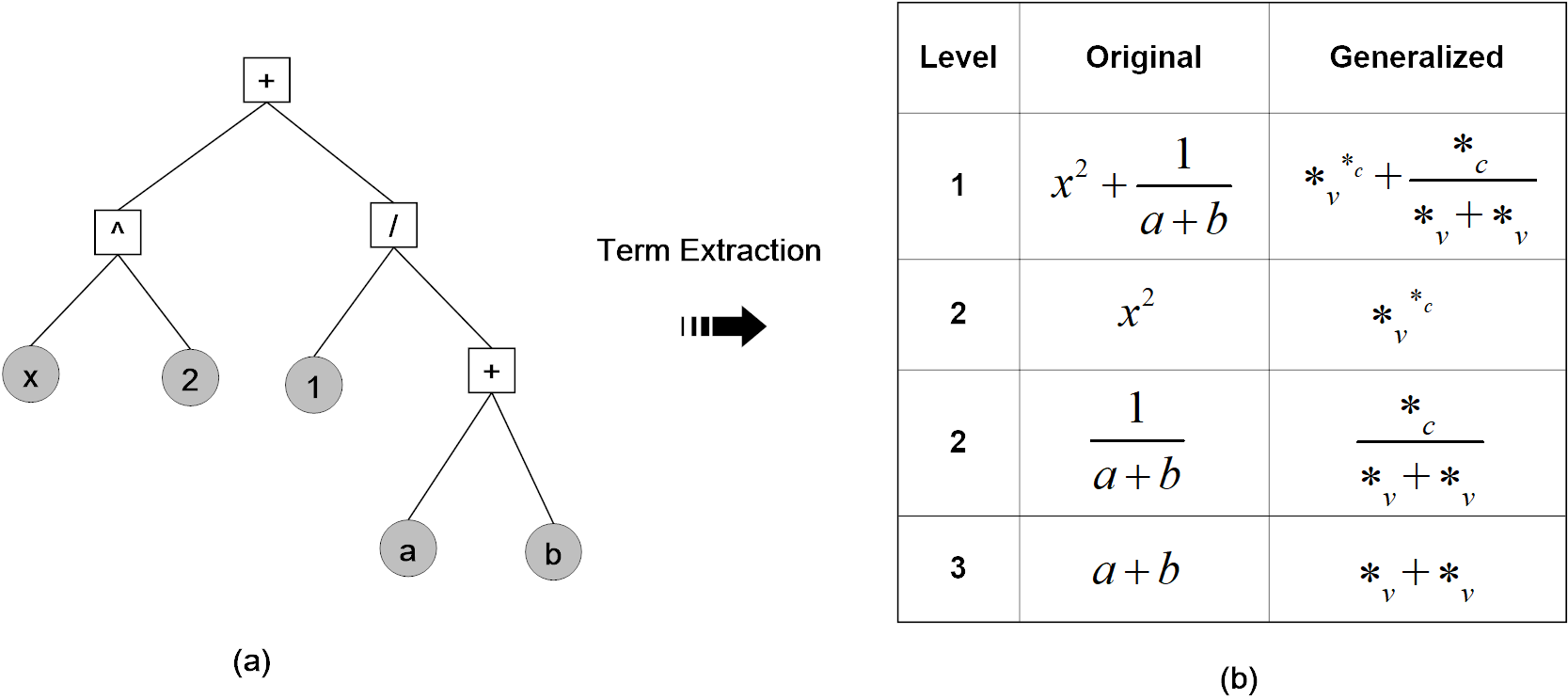}  
	\caption{(a) Semantic tree presentation and (b) term extraction of $x^{2}+\frac{1}{a+b}$}\vspace{-2ex}
	\label{fig:layoutexp}
\end{figure}

In this study, a formula semantic tree based approach is used for calculating $\lambda _{c}\left ( f \right )$. We first parse the \LaTeX{} expressions of formulae and convert them into Presentation MathML expressions. Then we construct a formula tree using a semantic tree-constructed algorithm proposed in \cite{lin2014mathematics} (Figure \ref{fig:layoutexp} (a) shows an example of semantic tree presentation of the formula $x^{2}+\frac{1}{a+b}$). After that, we extract formula terms hierarchically from the constructed semantic tree. The extraction algorithm is described in Algorithm \ref{alo:extraction}. In the proposed algorithm, there are two kinds of formula terms: original terms and generalized terms. The original terms are generated directly from the original substructures of the semantic tree presentation of the formula. The generalized terms are proposed by changing the variables and constants of the original terms into wildcards (describe the sketch of the formula structure, for fuzz representation). Variables are replaced by $*_{v}$, and constants are represented as $*_{c}$. There is a ``level'' attribute extracted for each term which denotes the level of the term in the semantic tree whose root's level is 1. Figure \ref{fig:layoutexp} (b) shows the sub-tree levels and terms of formula ``$x^{2}+\frac{1}{a+b}$''. The similar formula tree layout presentation has been proven an effective method for formula retrieval task \cite{gao2016math}.

\vspace{-1ex}

Based on the formula semantic tree and its extracted terms, the $\lambda _{c}\left ( f \right )$ can be calculated as:
$\lambda _{c}\left ( f \right ) = \sum_{i=1}^{N} \left ( L(t_{i}) \right )$, in which, $N$ is the total formula term number, $t_{i}$ is a formula term, and $1\leq i\leq N$, $L(t_{i})$ is the level of $t_{i}$ in the formula tree. The greater $\lambda _{c}\left ( f \right )$ is, the more complicated formula $f$ is.

\textbf{(c) Formula Evolution Probability Calculation}: Although Wikipedia provides multitudinous evidence that can be used to infer candidate evolution relations, some of them can be noisy. For instance, page \textit{``Artificial neural network''} has a link to page \textit{``Algorithm''}, but not every formulae in  ``Algorithm'' page can be evolutionary to the ``Activation function'' formula, some of them could be noisy. This kind of noisy formulae will not be useful for math-content understanding. To address this problem, we propose the formula evolution probability to characterize the reliability of a formula evolution relation. The evolution probability from $f_{a}$ to $f_{b}$ can be modeled as:
\begin{equation}
P(f_{a} \overset{e}{\rightarrow} f_{b}) = \pi \left ( \phi (f_{a}),\phi (f_{b}) \right )
\end{equation}
where, $P(f_{a} \overset{e}{\rightarrow} f_{b}) $ is the formula evolution probability from $f_{a}$ to $f_{b}$, $\phi$ is a representation function, which can project each formulae to a low-dimensional joint embedding space from context, layout and generality viewpoints. $\pi$ is probability scoring function based on the learned formula embeddings.

\begin{algorithm}[!htb]
\small
\caption{Formula Semantic Term Extraction}
\label{alo:extraction}
\renewcommand{\algorithmicrequire}{\textbf{Input:}}
\renewcommand{\algorithmicensure}{\textbf{Output:}}
\begin{algorithmic}[1]
\Require 
Formula Semantic Tree, $ST$
\Ensure 
Set of formula terms, $T$
\State Let $O(ST)$ be the original semantic tree
\State Let $G(ST)$ be the generalization of the semantic tree
\State Let $L(ST)$ be the level of the semantic tree
\Procedure{Extractor}{$ST, L(ST)$}
	\If {$ST$ is not a leaf}
		\State $T += ( O(ST),L(ST))$ \Comment{ original term}
		\State $T += ( G(ST),L(ST))$ \Comment{ generalized term}
		\For{$ST_{i} \leftarrow $ each child of $ST$}
			\State \Call{Extractor}{$ST_{i}, L(ST)+1$}	
		\EndFor
	\EndIf
\EndProcedure 
\end{algorithmic}
\end{algorithm}\vspace{-2ex}

In this work, we construct the formula representation via semi-supervised graphical learning. Following the evolution relations in FEM, we can simulate a random walk of fixed length $l$ with a set of parameters $\mathbf{\theta}$ to guide the walker on the graph. Let $f_{i}$ denote the $i_{th}$ formula in the walk, which can be generated by the following distribution:
\begin{equation}
\begin{matrix}
P(f_{i}|f_{i-1}) = tanh\left [ \tau ( P_{t}, P_{l} ,P_{g}| \mathbf{\theta} ) \right ]\\
= tanh\left [  \theta _{t}P_{t}(f_{i}|f_{i-1})+\theta _{l} P_{l}(f_{i}|f_{i-1})+ \theta _{g}P_{g}(f_{i}|f_{i-1}) \right ]
\end{matrix}
\end{equation}
where $P(f_{i}|f_{i-1})$ denotes the normalized transition probability between $f_{i}$ and $f_{i-1}$. $\tau(\cdot)$ is a trivariate function that can fusion three different kinds of  transition probabilities: context $P_{t}$, layout $P_{l}$, and generality $P_{g}$. $\mathbf{\theta} = \left \{ \theta _{t}, \theta_{l},\theta _{g} \right \}$ is the non-negative fusion parameters that control the contribution of each transition probability. For this study, we set $\theta _{t} =  \theta_{l} = \theta _{g} =1$, and more sophisticated parameter tuning will be saved for future. $tanh\left ( \cdot \right )$ is the hyperbolic tangent function for normalization.

For context transition probability estimation, a 250-word text window around the formulae was employed along with language model with Dirichlet smoothing \cite{zhai2001study}.
\begin{equation}
P_{t}\left ( f_{i}^{t} | f_{i-1}^{t} \right ) \propto P_{t}( f_{i-1}^{t} | f_{i}^{t})P_{t}(f_{i}^{t})
\label{equ:fcontext}
\end{equation}
$f_{\cdot}^{t}$ represents the context of a formula, $P_{t}\left ( f_{i}^{t} | f_{i-1}^{t} \right )$ is the posterior probability, $P_{t}( f_{i-1}^{t} | f_{i}^{t})$ is the $f_{i-1}^{t}$ likelihood given $f_{i}^{t}$, $p(f_{i}^{t})$ is assumed to be uniform. We hypothesize that if two formulae share the similar context, they may have a high evolution probability.

For formula layout, we calculate formula transition probability by leveraging formula semantic layout tree and its extracted terms:
\begin{equation}
P_{l}\left ( f_{i}^{t} | f_{i-1}^{t} \right ) = \frac{ \omega_{cov}(f_{i}^{l},f_{i-1}^{l}) \sum_{t_{n}\in f_{i-1}^{l}}[\omega_{gen}(t_{n}) \omega_{lel}(t_{n},f_{i}^{l},f_{i-1}^{l})]}{\sum_{t_{n}\in f_{i-1}^{l}}[\omega_{gen}(t_{n})]}
\label{equ:flayout} 
\end{equation}
where, $f_{\cdot}^{l}$ is the formula term set generated from the semantic layout tree; $\omega_{cov}(f_{i}^{l},f_{i-1}^{l}) = \frac{f_{i}^{l}\bigcap f_{i-1}^{l}}{|f_{i-1}^{l}|}$, denotes the ratio between matched term number and total term number of $f_{i-1}$; $\omega_{gen}(t_{n})$ is the penalty parameter for the generalized terms, if $t_{n}$ is a generalized term, $\omega_{gen}(t_{n})$ is empirically set to 0.5 \cite{lin2014mathematics}, otherwise, $\omega_{gen}(t_{n})=1$; $\omega_{lev}(t_{n},f_{i}^{l},f_{i-1}^{l})$ is the term level weight, affected by the minimum level distance of the matched formula term $t_{n}$ in $f_{i}$ and $f_{i-1}$, $\omega_{level}(t_{n},f_{i}^{l},f_{i-1}^{l}) = \frac{1}{1+min_{j}\left \{  |level(t,f_{i-1}^{l})-level_{j}(t,f_{i}^{l})| \right \}}$. We hypothesize that, if two formulae have similar layout trees, they may have a high evolution probability.

Formula generality transition probability can be calculated as:
\begin{equation}
P_{g} (f_{i}|f_{i-1}) = \lambda_{g} (f_{i-1})
\label{equ:fpopu}
\end{equation}
which is the formula generality of $f_{i-1}$. We hypothesize that, a fundamental formula can have a high evolution probability to its variants with more detailed contextual constraints.

We then use $\phi:F\rightarrow \mathbb{R}^{d}$ as the mapping function (from formula vertexes) for representation learning, where $d$ specifies the number of dimensions. $\phi$ is a matrix of size $\left | F \right | \times d$ parameters. Feature learning methods are based on the Skip-gram architecture \cite{mikolov2013distributed}. For every starting formula vertex $f \in F$, we define $N_{S}(f) \subset F$ as a network neighborhood (``context'') of vertex $f$ generated through the proposed neighborhood sampling strategy $S$ (semi-supervised random walk guided by $\mathbf{\theta}$). The objective function can be formalized as:
\begin{equation}\label{equ:max}
\underset{\phi}{max}\sum_{f\in F} log P(N_{S}(f)|\overrightarrow{\phi(f)})
\end{equation}
Stochastic gradient ascent is used to optimize the joint embedding model parameters of $\overrightarrow{\phi(\cdot)}$. Negative sampling \cite{mikolov2013distributed} is applied for optimization efficiency. In this study, we use cosine similarity (with a ReLU function) of the optimized formula representations as $\pi$ for scoring the formula evolution probability.

By calculating the evolution probability for formula relation, we are able to rule out the noisy relations (with low evolution probabilities) and better explore the formula evolution trajectory to help users understand the essence of the target formula.

\begin{figure*}[h]\centering
	\includegraphics[width=1.7\columnwidth]{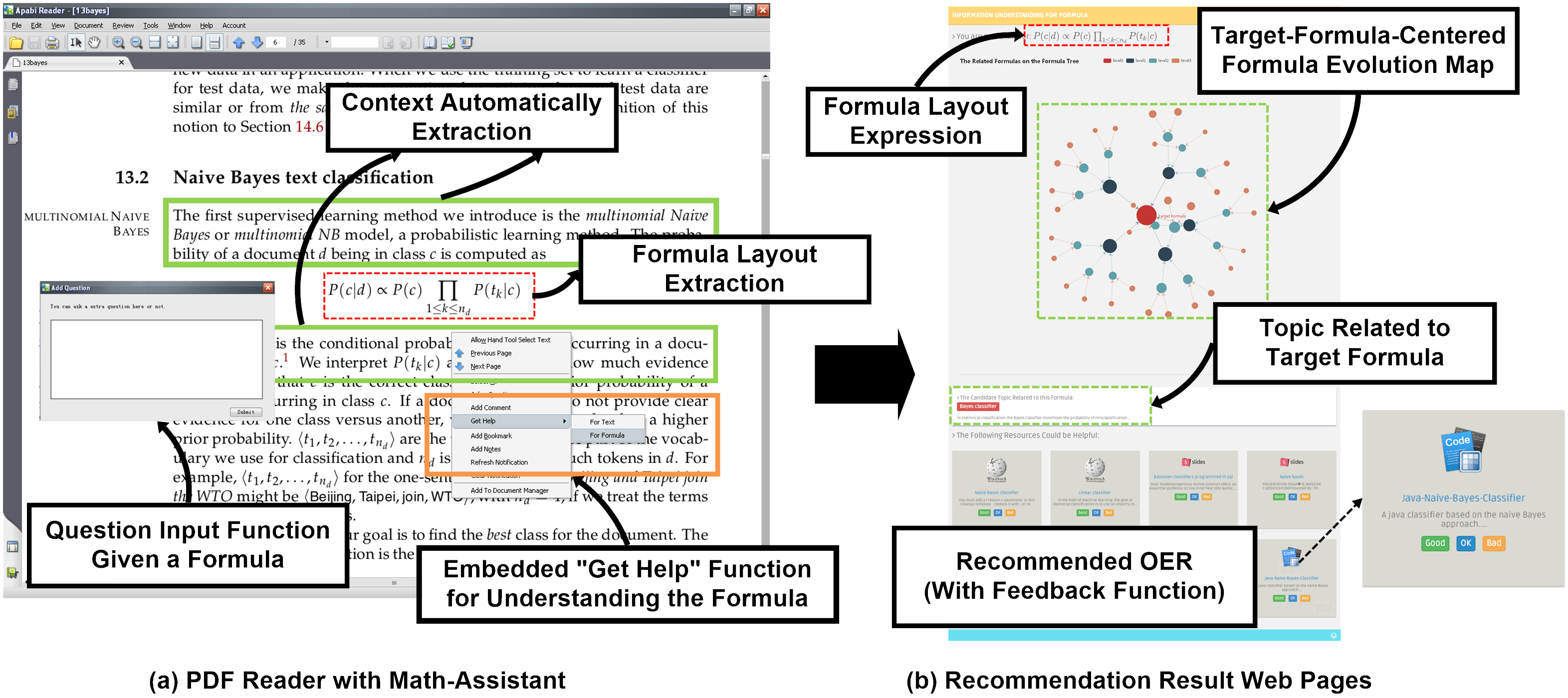}  
	\vspace{-2ex}
	\caption{PDF Reader with Math-Assistant (PRMA) System }
	\label{fig:system}
 \end{figure*}
 
\vspace{-2ex}\subsection{Math-Information Need Characterization}\label{sec:minfor}

To help students better understand the math-content of a scientific publication in a course environment, we design a novel system, PDF Reader with Math-Assistant (PRMA). As Figure \ref{fig:system} shows, the new system has three main functions:

$\bullet$ Capture evidence and characterize students' emerging implicit/explicit information needs when encounter a formula understanding problem. For instance, students can ask a specific question given a formula (explicit information need), or easily highlight a formula with mouse in the paper, as evidence of an implicit information need. In either case, the PRMA is able to extract the formula layout presentation and formula context from the target PDF paper.

$\bullet$ Automatically project the target formula onto the formula evolution map, which allows students to navigate the formula evolution trajectory in FEM while helping them understand the target formula.

$\bullet$ Automatically recommend high quality OERs for the target formula, such as video lectures, slides, source code, or Wikipedia pages, to resolve students' information needs while helping them understand the formula. (We crawl and pre-index massive OERs by using meta-search algorithms provided in \cite{liu2015scientific}.) 


At the front-end, the mathematical expressions are obtained by a symbol dominance based formulae recognition algorithm proposed in \cite{ZhangGYLJT17}, and the formula context is extracted by a PDF parser. At the backend, the PRMA has access to the generated FEM, the list of assigned class readings (title, abstract, full content, associated topics and citation information), and the formula-based OER recommendations algorithm.

\begin{table}[!htb]
\centering
\footnotesize
\caption{Formula projecting features for math-information need characterization*} \vspace{-2ex}
\label{tab:fpf}
\begin{threeparttable}
\begin{tabular}{l  l  p{2.5cm} l}
\hline
No. & &Projecting feature & Mathematical Definition \\\hline
1&\multicolumn{1}{>{\columncolor{red}}l} {~}& Formula Context Feature & $ p( f_{?}^{t} | f_{c}^{t})p(f_{c}^{t})$ \\ 

2&\multicolumn{1}{>{\columncolor{red}}l} {~}& Formula Context Keyword Feature & $ \sum _{k_{i} \in f_{?}^{t}} p( f_{?}^{k_{i}} | f_{c}^{t})p(f_{c}^{t})$  \\ 

3&\multicolumn{1}{>{\columncolor{red}}l} {~}& Question Text Feature &$ p( f_{?}^{q_{t}} | f_{c}^{t})p(f_{c}^{t})$ \\

4&\multicolumn{1}{>{\columncolor{red}}l} {~}& Question Text Keyword Feature & $ \sum _{q_{k,i} \in f_{?}^{q_{t}}} p( f_{?}^{q_{k,i}} | f_{c}^{t})p(f_{c}^{t})$\\

5&\multicolumn{1}{>{\columncolor{yellow}}l} {~}&Formula Layout Feature & $\frac{ \sum_{t_{i}\in f_{?}^{l}}[\omega_{gen}(t_{i}) \omega_{lel}(t_{i},f_{c}^{l},f_{?}^{l})]}{[\omega_{cov}(f_{c}^{l},f_{?}^{l})]^{-1}\sum_{t_{i}\in f_{?}^{l}}[\omega_{gen}(t_{i})]}$ \\ 

6&\multicolumn{1}{>{\columncolor{blue}}l} {~}&Paper Idea Feature & $p( f_{?}^{p_{abs}} | f_{c}^{t})p(f_{c}^{t})$  \\ 

7&\multicolumn{1}{>{\columncolor{blue}}l} {~}&Paper Keywords Feature &$ \sum _{k_{i} \in f_{?}^{p_{K}}} p( f_{?}^{k_{i}} | f_{c}^{t})p(f_{c}^{t})$  \\ 

8&\multicolumn{1}{>{\columncolor{blue}}l} {~}&Weekly Topic Feature & $\sum _{w_{i} \in f_{?}^{p_{W}}} p( f_{?}^{w_{i}} | f_{c}^{t})p(f_{c}^{t})$ \\ 

9&\multicolumn{1}{>{\columncolor{green}}l} {~}&Context Evolution Feature & $p( f_{m}^{t} | f_{c}^{t})p(f_{c}^{t})$ \\ 

10&\multicolumn{1}{>{\columncolor{green}}l} {~}&Layout Evolution Feature & $\frac{ \sum_{t_{i}\in f_{m}^{l}}[\omega_{gen}(t_{i}) \omega_{lel}(t_{i},f_{c}^{l},f_{m}^{l})]}{[\omega_{cov}(f_{c}^{l},f_{m}^{l})]^{-1}\sum_{t_{i}\in f_{m}^{l}}[\omega_{gen}(t_{i})]}$  \\ 

11&\multicolumn{1}{>{\columncolor{green}}l} {~}&Generality Evolution Feature & $\lambda _{g} (f_{c})$\\ 

12&\multicolumn{1}{>{\columncolor{green}}l} {~}& Evolution Distance Feature & $|f_{m}\rightsquigarrow f_{c}|$  \\ \hline
\end{tabular}
\begin{tablenotes}
    \footnotesize
    \item \colorbox{red}{{\color{red}1}} Formula Text Feature Group~~ \colorbox{yellow}{{\color{yellow}2}} Formula Layout Feature Group~~ \item \colorbox{blue}{{\color{blue}1}} Paper Content Feature Group~~\colorbox{green}{{\color{green}2}} Formula Evolution Feature Group
    \item *Because of the space limitation, the detailed feature description(hypothesis) will be available in https://github.com/GraphEmbedding/FEM
    \end{tablenotes}
\end{threeparttable}
\end{table}

The algorithms presented in the next section can recommend the optimized OERs given the math-information need, which will be able to help readers better understand the essence of the targeted formula. Meanwhile, readers can also provide usefulness feedback for system recommended OERs. For instance, as Figure \ref{fig:system} shows, readers can click \textit{``Good'', ``OK''}, or \textit{``Bad''} for each recommended OER given their information needs. The judgments and student click information will be saved as system logs. The formula evolution explore behavior (click the formula evolution map) will also be recorded, which will be important for formula-based OER recommendation algorithms, i.e., training learning to rank model, and algorithm evaluation. 


Although FEM can be potentially helpful for MCU, it's still challenging to characterize the student's emerging information needs while facing a formula in the target paper. The critical problem is how to ``project'' a puzzling formula to one or a number of formula vertex(es) in the FEM.


To address this problem, by using PRMA, we employ multiple Formula Projecting Features (FPF) to characterize the math-information need. The proposed FPF mainly focuses on four aspects: (1) the math-information need can be related to formula layout, (2) related to the text information of formula, e.g., students' questions about the formula, context and their associated topics, (3) related to the content of paper where formula exists, i.e., abstract, keywords and weekly topics in the syllabus (in a course environment), (4) related to the formulae located in the evolution trajectories in FEM given a matched formula. Detailed math definition is provided in Table \ref{tab:fpf}.

\vspace{-2ex}\subsection{OER Recommendation via FEM Mining} \label{sec:OERrec}

\begin{figure}[!htb]\centering
	\includegraphics[width=0.9\columnwidth]{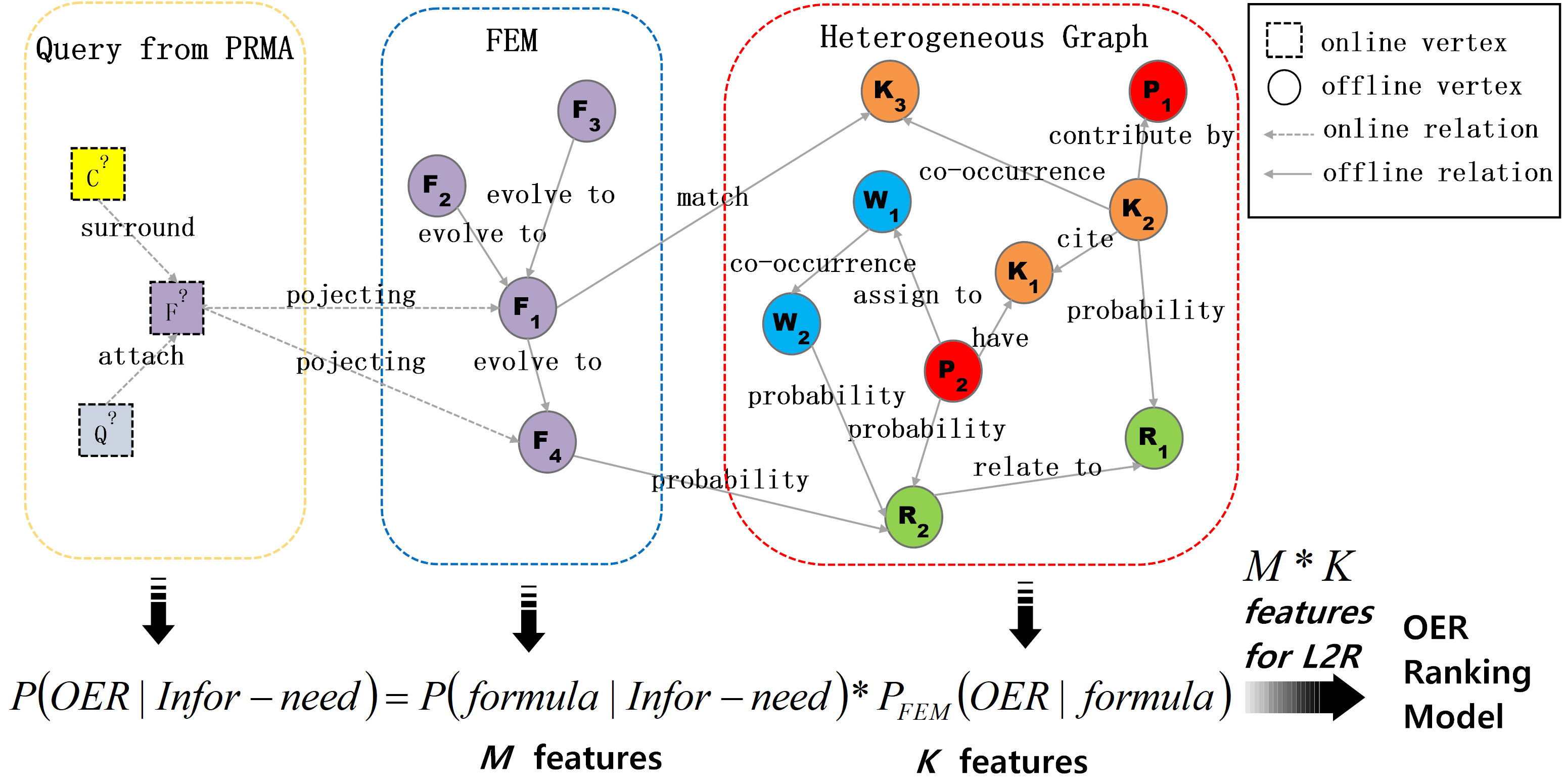}
	\vspace{-2ex}
	\caption{Graphical representations of OER recommendation via FEM mining}
	\label{fig:oer}
\end{figure}

Figure~\ref{fig:oer} illustrates the graphical OER recommendation via FEM mining towards the mathematical query from PRMA, the vertexes and edges are depicted in Table~\ref{tab:graph}. There are totally 48 offline OER ranking features (ORF) constructed\footnote{Because of the space limitation, we cannot provide more detailed OER ranking features. The detailed feature list will be available in https://github.com/GraphEmbedding/FEM} for this study, which can be divided into two groups:

(1) text-mining-based feature group, for instance, we calculate the $p(OER|formula)$ based on the OER's text description and formula context using the language model. Note that we utilize the reading paper's content information (abstract, keywords, and weekly topics) for constructing text-mining-based features, which means even for a same formula, if the reading paper is changed, the recommended OERs will change correspondingly. 

(2) Heterogeneous-graph-ranking feature group, a formula vertex in FEM can random walk to the OERs in the scholarly heterogeneous graph, i.e., OER ranking given a formula vertex in FEM. In this study, we employ meta-path plus random walk from the formula to candidate OER as graph-ranking features. For instance, $F^{*} \overset{m}{\rightarrow} K \overset{p}{\rightarrow} R^{?}$ is a graph-ranking feature, which denotes that if an OER has a high probability relation from the keywords (topics) vertexes extracted from question formula's context, this OER should be recommended. The random walk probability can be estimated by:
\begin{equation}
r(v_i^{(1)},v_j^{(l+1)}) = \sum_{t= v_i^{(1)}\rightsquigarrow v_j^{(l+1)}} RW(t), \ RW(t) = \prod_j w(v_{ij}^{(j)},v_{i,j+1}^{(j+1)})
\end{equation}
where $t$ is a tour from $v_i^{(1)}$ to $v_j^{(l+1)}$ following the meta-path\cite{Sun+11}, and $RW(t)$ is the simulated random walk probability of the tour $t$. Suppose $t = (v_{i1}^{(1)}, v_{i2}^{(2)}, \ldots, v_{i{l+1}}^{(l+1)})$, $w(v_{ij}^{(j)},v_{i,j+1}^{(j+1)})$ is the weight of edge $v_{ij}^{(j)} \rightarrow v_{i,j+1}^{(j+1)}$. More detailed algorithm can be found in \cite{liu2015scientific}.

In section \ref{sec:minfor}, we proposed a number of FPF (i.e., $M$ formula projecting features, see Table \ref{tab:fpf}) for formula projecting, and the ORF (i.e., $K$ OER ranking features) are presented in this section. By using the formulae in FEM as the transition, we integrate FPF and ORF to recommend useful OER for math content understanding. The OER recommendation probability $P(v_{r}|v_{f^{?}})$, i.e., an OER $r$ based on the query formula $f^{?}$, can be calculated as:
\begin{equation}
P(v_{r}|v_{f^{?}}) = \sum_{m=1}^{M} \sum_{k=1}^{K}\omega_{m,k}\cdot \mathbb{E}_{F_{m} \in FPF } (v_{f^{?}},v_{f}) \cdot \mathbb{E}_{F_{k} \in ORF } (v_{f},v_{r})
\end{equation}
Here, $\omega_{m,k}$ is the feature weight for $m_{th}$ FPF and $k_{th}$ ORF, $\mathbb{E}_{F_{m} \in  FPF } \\(v_{f^{?}} , v_{f})$ is the online query formula projecting probability based on different FPF, $\mathbb{E}_{F_{k} \in  ORF} (v_{f},v_{r})$ is the OER ranking probability based on different ORF. There are totally $M*K$ features for OER recommendation. In order to avoid laborious parameter tuning, we utilize learning to rank algorithm \cite{li2014learning} to jointly optimize the feature weights for FPF and ORF. The training data (OER usefulness judgments) are collected via PRMA system.

\begin{table}[htb]
\footnotesize
\centering
\caption{Vertexes and edges of OER recommendation via FEM mining} \vspace{-2ex}
\label{tab:graph}
\begin{tabular}{l  p{6.5cm} }
\toprule
\textbf{Vertex}       & \textbf{Description}                                        \\ \midrule
$R$           &  Open Education Resource             \\ 
$P$           &  Paper                                              \\ 
$K$           &  Keyword      \\ 
$W$           &  Weekly Topic (from Syllabus)            \\ 
$F$           &  Formula           \\ 
$F^{?}$       &  Query Formula     \\ 
$C^{?}$       &  Context of Query Formula   \\
$Q^{?}$       &  User Additional Question   \\ \midrule
\textbf{Edge}       & \textbf{Description}                                        \\ \midrule
$P \overset{h}{\rightarrow} K$           &  Paper is related to keyword (using Labeled LDA \cite{ramage-LLDA})            \\ 
$P \overset{a}{\rightarrow} W$           &  Paper is assigned to weekly topic (probability)                     \\ 
$P \overset{p}{\rightarrow} R$           &  Paper-resource relationship based on  $p(R|P)$ (language model)       \\ 
$K \overset{cite}{\rightarrow} K$           &  Keyword cites keyword (probability)                                          \\ 
$K \overset{co}{\rightarrow} K$           &  Keyword-keyword co-occurrence (probability)                             \\ 
$K \overset{cont}{\rightarrow} P$           &  Keyword is contributed by paper (using PageRank with prior \cite{white-PRprior})                            \\ 
$K \overset{p}{\rightarrow} R$           &  Keyword-resource relationship based on  $p(R|K)$ (language model)      \\ 
$W \overset{co}{\rightarrow} W$           &  Weekly topic-weekly topic co-occurrence (probability)              \\ 
$W \overset{p}{\rightarrow} R$           &  Weekly topic-resource relationship based on $p(R|W)$ (language model) \\ 
$R \overset{r}{\rightarrow} R$           &  OER is related to OER (collected from service sites) \\ 
$F \overset{p}{\rightarrow} R$           &  Formula context-OER content relation based on  $p(R|F)$ (language model)         \\ 
$F \overset{m}{\rightarrow} K$           &  Formula context is related to keyword (greedy match algorithm)    \\ 
$F \overset{e}{\rightarrow} F$           &  Formula evolves to formula (probability)   \\ 
$C^{?} \overset{s}{\rightarrow} F^{?}$           &  Context surrounds formula \\
$Q^{?} \overset{at}{\rightarrow} F^{?}$           &  Additional question is attached to formula   \\
$F^{?} \overset{p}{\rightarrow} F$           &  Query formula is online  projected to the formula on FEM   \\ \bottomrule
\end{tabular} 
\end{table}
\vspace{-1ex}\section{Experiment}

\subsection{Dataset and Experiment Setting}

We tested this reading system and the associated mathematics content understanding algorithms in a real learning environment. Two graduate-level information retrieval courses at Indiana University were used for this experiment. A total of 52 students (Masters and Ph.D.s) voluntarily participated this experiment, and they were required to use the PRMA system for eight weeks (with 15 required readings, 10 chapters of an IR book, and 5 ACM journal papers). They could use PRMA with university account, and PRMA enables formula understanding function (e.g., highlight a formula in the reading and access formula evolution trajectories in FEM as well as recommended OERs). Meanwhile, we asked each participant to provide OER relevance judgments for the system-recommended OERs. There were a total of 7,099 valid judgments collected (for 622 student requests), and we used those judgments to train the learning to rank model and to evaluate the algorithm performance. Among the 622 student requests, only 29 (4.7\%) requests contained an explicit question. This phenomenon indicates that most students don't want to input a specific question when facing a formula in PRMA.

At the backend of PRMA, we created a heterogeneous graph for OER recommendation. For paper vertexes, we used 248,893 publications from 1,553 venues (in ACM digital library). The paper vertexes were connected to 7,190 keyword labeled topics, and the publication data were also used to generate formula birth time $\lambda _{t}\left ( f \right )$. According to the syllabus, there were a total of 60 weekly topics. By using meta-search, we collected a total of 1,112,718 OERs.

We used a Wikipedia dump of July 30, 2014. There were 358,116 raw formulae, 34,683 formula home pages, 198,336 page-formula ownership relations, and 74,947,670 page hyperlinks. For the experiment, we kept the formulae featuring at least two variables and three operators. There were 194,150 formulae left, and the generated FEM had 21,292,157 potential formula evolution relations. 

As Figure \ref{fig:system} (b) shows, for each query formula, we visualized a three-level (distance) target-formula-centered evolution sub-graph. Meanwhile, the visualized formula vertex had at least 0.5 probability ($P(f_{a} \overset{e}{\rightarrow} f_{b}) \geq $ 0.5) to connect to its neighbor.

\vspace{-2ex}\subsection{Experiment Results}

\begin{figure}[!htb]\centering\vspace{-2ex}
	\includegraphics[width=1.0\columnwidth]{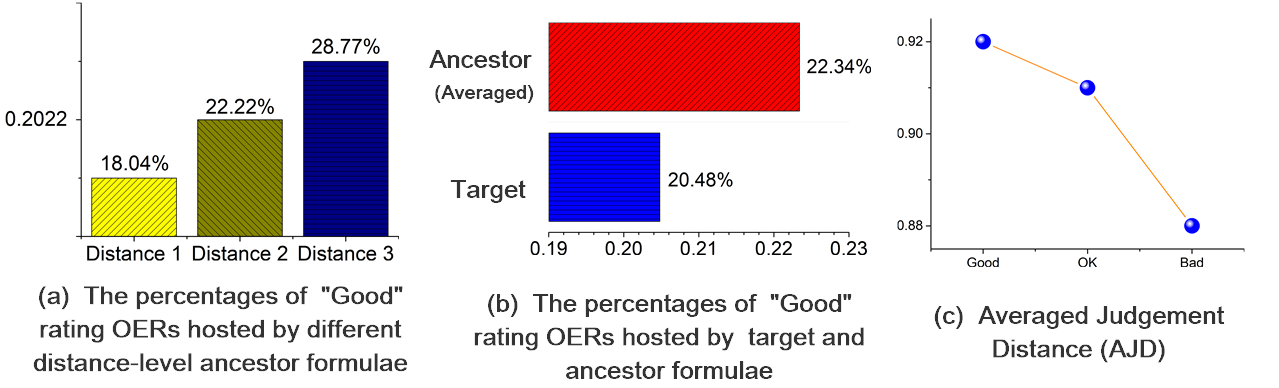} 
	 \vspace{-2ex}
	 \caption{Statistics of OER judgements}
	\label{fig:judge}\vspace{-2ex}
 \end{figure}

Among the 622 student requests, 610 of them (98.1\%) contained at least one OER rated as ``Good'' or ``OK'' for their ``mathematics content understanding''. For all the OER judgments, participants rated 19.72\% of the recommended OERs as \textit{``Good''}, 37.61\% as \textit{``OK''}, and 42.67\% as \textit{``Bad''}. Note that, the students were asked to rate at least top 5 OERs for each request, because we need to collect a amount of \textit{``Bad''} judgments for model training and evaluation. Based on the target-centered formula evolution map provided by PRMA, users could not only consume the recommended OERs for target formula but also freely explore the formula evolution map to check the recommended OERs for ancestor formulae. For any formula request, PRMA could record the user OER judgement, the ancestor formula (on FEM) that hosted the OERs, and the distance from ancestor formula to target formula on FEM. As Figure \ref{fig:judge} (a) shows, while the (evolutionary) distance between target and ancestor formula increasing, the percentage of ``Good'' judgements for OERs (ancestor formula hosted) raises (\textit{from 18.04\% to 28.77\%}). Meanwhile, as Figure \ref{fig:judge} (b) indicates, there are more percentages of ``Good'' rating OERs from the ancestor formulae. This finding demonstrates that the ancestor formulae on FEM can be especially important to assist students better consume the math-content in a paper, and, when the target formula is complex, the background information (e.g., the ancestor formulae on FEM) can be more useful.

In order to further explore the relationship between OER judgements and evolutionary distance, we calculate the Average Judgement Distance (AJD) from ancestor to question (target) formula on FEM for each type of judgements: $AJD_{type} = \frac{\sum_{i=1}^{N_{type}} d_{i}}{N_{type}}$. $N_{type}$ is the total number of a specific type of judgement (e.g., ``Good'', ``OK'', etc.); $d_{i}$ should be one of $\left \{ 0,1,2,3 \right \}$, representing the evolutionary distance between OER hosted formula and target formula, for $i_{th}$ judgement of this specific type. The result, $AJD_{``Good''}$ (0.92) $>$ $AJD_{``OK''}$ (0.91) $>$ $AJD_{``Bad''}$ (0.88), indicates the remoter ancestor formulae (with recommended OERs) can be more helpful for math information understanding. This finding also proves FEM can be effective to address the student's math-information need for math-understanding (without FEM, it is not feasible to trace the ancestor formulae and math-evolution for information understanding). 

\vspace{-2ex}\begin{table}[htbp]
\footnotesize
\centering
\caption{Baseline groups and comparison groups for OER ranking experiment} \vspace{-2ex}
\label{tab:expgroup}
\begin{threeparttable}
\begin{tabular}{ m{0.05cm}  m{1.5cm}  m{5.7cm} }
\toprule
\multicolumn{3}{l}{\textbf{Baselines}}                                    \\ \midrule
\multicolumn{1}{>{\columncolor{red}}l} {~} & $Rank_{abs}$           &  Use the reading's abstract to represent the user's math-information needs, then generate the OER ranking based on $p(OER|abstract)$ (using Language Model with Dirichlet prior smoothing \cite{zhai2001study}, the same below).              \\ 
\multicolumn{1}{>{\columncolor{red}}l} {~} & $Rank_{keyword}$           & Use the reading's keywords to represent the user's math-information needs: $p(OER|keywords)$.                                              \\ 
\multicolumn{1}{>{\columncolor{red}}l} {~} & $Rank_{context}$           &  Use the formula's context to represent the user's math-information needs: $p(OER|context)$.      \\ \midrule
\multicolumn{3}{l}{\textbf{Comparison Groups}}                                 \\ \midrule
\multicolumn{1}{>{\columncolor{yellow}}l} {~} & $L2R_{layout}$           &  L2R (learning to rank) model using formula layout feature\cite{lin2014mathematics} plus all OER ranking features.            \\ 
\multicolumn{1}{>{\columncolor{yellow}}l} {~} &$L2R_{text}$           &  L2R model using formula context feature, formula context keyword feature, question text feature and question text keyword feature plus all OER ranking features.       \\ 
\multicolumn{1}{>{\columncolor{yellow}}l} {~} &$L2R_{content}$           &  L2R model using paper idea feature, paper keywords feature and weekly topic feature plus all OER ranking features.      \\ 
\multicolumn{1}{>{\columncolor{yellow}}l} {~} &$L2R_{multiple}$           &  L2R model using formula layout feature group, formula text feature group and paper content feature group plus all OER ranking features.                                         \\ 
\multicolumn{1}{>{\columncolor{blue}}l} {~} &$L2R_{all}$           &  L2R model using all  formula projecting features plus all OER ranking features.                            \\ \bottomrule
\end{tabular}
\begin{tablenotes}
    \footnotesize
    \item \colorbox{red}{{\color{red}1}} Without FEM \quad \colorbox{yellow}{{\color{yellow}2}} With Partial FEM \quad
    \colorbox{blue}{{\color{blue}3}} With Complete FEM
    \end{tablenotes}
\end{threeparttable}
\end{table}\vspace{-2ex}

As this study is not focusing on learning to rank (L2R), we used a relative simple list-wise algorithm, Coordinate Ascent\cite{metzler2007linear}, which iteratively optimizes a multivariate objective ranking function, for formula and OER features integration and algorithm evaluation. Meanwhile, we needed to employ the baseline groups for comparison. However, as mathematics content understanding is a newly proposed problem, and few existing algorithms addressed this problem. We chose three classic methods as baseline groups (without FEM assisted) and five L2R models for different feature groups (with partial or complete FEM assisted) as comparison groups. The baselines (e.g., text and formula retrieval models) and comparison groups are listed in Table \ref{tab:expgroup}.

\begin{table*}[bhtp]
\footnotesize
\centering
\caption{Measures of different OER ranking algorithms (Significant test: $\mathbf{L2R_{all}}$ vs. other groups; \textsuperscript{\dag}$\mathbf{p < 0.01}$, \textsuperscript{\dag\dag}$\mathbf{p < 0.001}$, \textsuperscript{\dag\dag\dag}$\mathbf{p < 0.0001}$)}\vspace{-2ex}
\label{tab:ranking}
\begin{threeparttable}
\begin{tabular}{l p{2cm}  p{1.5cm} p{1.5cm} p{1.8cm} p{1.5cm} p{1.5cm} p{1.5cm} l}
\hline
& \textbf{Ranking} & \textbf{NDCG@3} & \textbf{NDCG@5} & \textbf{NDCG@all} & \textbf{P@3} & \textbf{P@5} & \textbf{MAP} & \textbf{MRR} \\ \hline
\multicolumn{1}{>{\columncolor{red}}l} {~}& $Rank_{abs}$ &  0.5536 & 0.5860 & 0.7156 & 0.5544 & 0.4932 & 0.7309 & 0.7764  \\ 
\multicolumn{1}{>{\columncolor{red}}l} {~}& $Rank_{keyword}$  & 0.5744 & 0.6004 & 0.7268 & 0.5635 & 0.4973 & 0.7344 & 0.7833  \\ 
\multicolumn{1}{>{\columncolor{red}}l} {~}& $Rank_{context}$ & 0.6483 & 0.6642 & 0.7589 & 0.6293 & 0.5422 & 0.7689 & 0.8328  \\ 
\multicolumn{1}{>{\columncolor{yellow}}l} {~}& $L2R_{layout}$ & 0.5771 & 0.6031 & 0.7253 & 0.5680 & 0.5000 & 0.7393 & 0.7861  \\ 
\multicolumn{1}{>{\columncolor{yellow}}l} {~} &$L2R_{text}$ & 0.6408 & 0.6531 & 0.7567 & 0.6202 & 0.5361 & 0.7671 & 0.8509  \\ 
\multicolumn{1}{>{\columncolor{yellow}}l} {~} &$L2R_{content}$ & 0.6571 & 0.6731 & 0.7651 & 0.6440 & 0.5551 & 0.7815 & 0.8497 \\ 
\multicolumn{1}{>{\columncolor{yellow}}l} {~} &$L2R_{multiple}$ & 0.6798 & 0.6908 & 0.7777 & 0.6497 & 0.5612 & 0.7961 & 0.8717  \\ 
\multicolumn{1}{>{\columncolor{blue}}l} {~} & $\boldsymbol{\mathit{L2R_{all}}}$   & \textit{\textbf{0.7150}}\textsuperscript{\dag\dag\dag} & \textit{\textbf{0.7252}}\textsuperscript{\dag\dag\dag} & \textit{\textbf{0.7969}}\textsuperscript{\dag\dag\dag} & \textit{\textbf{0.6712}}\textsuperscript{\dag\dag} & \textit{\textbf{0.5776}}\textsuperscript{\dag\dag} & \textit{\textbf{0.8171}}\textsuperscript{\dag\dag\dag} & \textit{\textbf{0.8848}}\textsuperscript{\dag}  \\ \hline
\end{tabular}  
\begin{tablenotes}
    \footnotesize
    \item \colorbox{red}{{\color{red}1}} Without FEM Assisted \qquad \quad \colorbox{yellow}{{\color{yellow}2}} With Partial FEM Assisted \qquad \quad \colorbox{blue}{{\color{blue}3}} With Complete FEM Assisted
    \end{tablenotes}
\end{threeparttable}
\end{table*}

From an NDCG viewpoint, we scored $Good = 2, OK= 1$, and $Bad= 0$. The OER recommendation performance can be found in Table~\ref{tab:ranking}. As the OER recommendation was more like a QA problem and students were more interested to find the first useful resource, we used MRR (Mean Reciprocal Rank) as the metric to train the learning to rank model. For evaluation, 10-fold cross-validation was used. 

From a performance viewpoint, evaluation results show that, first, FEM can provide important information for OER recommendation and math-understanding. For instance, $L2R_{all}$ (the best performed method empowered with complete FEM information), compared with the baseline groups (without FEM assisted), P@3 has an average increase of 15.2\%, NDCG@3 improves with 20.8\%, MAP increases 9.7\% and MRR enhances 10.9\%. Meanwhile, MRR score of $L2R_{all}$ is higher than 0.88 (which means students are finding the useful OERs (``Good'' or ``OK'') in the 1st position in almost result ranking list). This finding is also confirmed in the exit survey where 72.73\% of participants believe the PRMA system along with recommended OERs can provide precise and useful information for math-understanding. It is clear that FEM plays a critical role in the proposed framework, and FEM can provide very helpful information for math content understanding. 

Second, based on the student judgments, we found that a number of formula projecting features, can be potentially useful. From a ranking viewpoint, all L2R models outperform the baseline groups of $Rank_{abs}$ and $Rank_{keyword}$ (L2R model using only formula layout feature group or formula text feature group can not outperform the baseline of $Rank_{context}$). By combining multiple formula projecting feature groups, the ranking model $L2R_{multiple}$ outperforms all baseline groups (including $Rank_{context}$), which also proves L2R approach is an effective method for integrating features.

Third, while evolution relations among massive formulae are very useful for math-understanding, various kinds of information, i.e., formula layout, context, and generality, can be all useful for evolution relation discovery. For instance, though the comparison groups (based on partial FEM features) perform decently in the experiment, $L2R_{all}$ (with comprehensive FEM mining features) is significantly superior ($p<0.0001$) than all other groups for almost all the evaluation metrics. This finding supports our initial hypothesis that evolution relation is a latent variable hiding behind formula context and layout, and we can hardly explore it by using a single kind of evidence.


\vspace{-2ex}\subsection{Exit Survey}
The goal of this proposed study is to design a novel algorithm/system to assist students to better understand the mathematics content in a paper. Although from an OER ranking viewpoint the experiment results are positive, we designed another exit survey to further proof the usefulness of the new reading environment and the effectiveness of the new formula-understanding method. In the survey, we asked each participant seven questions (at the end of the experiment), including \textbf{``precision''}(Q1), \textbf{``satisfaction''} (Q2),  \textbf{``usefullness''} (Q3), \textbf{``relevance''} (Q4),  \textbf{``user-friendliness''} (Q5), \textbf{``usability''} (Q6) and \textbf{``effectivity''} (Q7)\footnote{More detailed information can be found at https://github.com/GraphEmbedding/FEM}. Based on the students' feedback, 72.73\% of participants believed the new system along with the formula-understanding method can provide precise and useful information, and 43.75\% find the proposed method can help them better understand the math-content in a paper ``most of the time" (another 31.25\% reported it is helpful ``about half of the time"). From a system usability perspective, 78.85\% of participants found the formula highlight function in PDF and OER recommendation functions were easy or very easy to use.

Overall, 63.63\% of participants reported that the system and formula understanding method can be helpful or very helpful to assist them to better understand the target paper (only 12.12\% reported the new functions are not helpful, and others were neutral). For the OER usefulness, 75.75\% participants are satisfied with the quality of the recommended OERs (especially for videos and slides). Participants reported, comparing narrative content, the OERs are more helpful for math-understanding. It is clear that the proposed system and math information understanding method achieves the goals, and the algorithms developed in this paper can be promising for scaffolding in education domain.

\vspace{-2ex}\section{RELATED WORK}

\textbf{Scientific Information Understanding:} Help students better understand and consume publications is an essential task in education and cyberlearning domains. Since 1976~\cite{wood1976role}, the term ``scaffolding'' has been widely used in educational research~\cite{pea2004social, puntambekar2005tools}. In particular, the concept of scaffolding is applied to the studies of computer-assisted learning environments, also known as computer-mediated scaffolding \cite{puntambekar2005tools}. One of the most recent efforts utilizes existing social tagging and annotation tools. Social tagging/annotation has produced positive results in a number of tasks, including the promotion of learning~\cite{johnson2010individual, su2010web, wolfe2008annotations}. Prior studies, however, also found those scaffolding approaches, by leveraging social tagging, can be quite limited~\cite{novak2012educational, liu2015scientific}, and students cannot essentially benefit from such systems when reading a challenging text. Researchers only recently began to focus on the usefulness of Open Educational Resources (OERs). For instance, Dennis, et al.~\cite{dennisimproving} found that an additional video presentation had significant positive impacts on students' learning, and more recently, Liu~\cite{liu2013generating, liu2013answering} found ODRs, e.g., presentation videos/slides and Wikipedia pages, can help scholars better understand scientific readings. Meanwhile, more recent studies~\cite{liu2015scientific, SIGIR-2016-OER} found that text and graph mining methods can be used to automatically recommend high quality OER to help students understand the paper text content. However, this approach cannot be applied to address the formula understanding problem. 

\textbf{Scientific Topic Evolution:} Topic dynamics and evolution has been recently investigated. Laura Dietz et al.~\cite{dietz2007unsupervised}, for example, devised a probabilistic topic model that explains the generation of documents. The model incorporated topical innovation and topical inheritance via citations. Blei and Lafferty~\cite{blei2006dynamic} proposed a Dynamic Topic Model (DTM), which explicitly characterized the chronological nature of sequential corpora by utilizing a Markov chain of term distributions over time. Based on \cite{blei2006dynamic}, Gerrish and Blei proposed the Document Influence Model (DIM)~\cite{gerrish2010language}. This model respected the ordering of the documents and not only tracked how underlying theme has changed over time, but also captured how past articles exhibit varying influence on future articles. More recently, Jiang et al.~\cite{jiang2015chronological} investigated topic evolution problem by integrating both text and citation data. Unlike earlier efforts, \cite{jiang2015chronological} generated a heterogeneous graph with various relations between topics and paper, i.e., citation and topic evolution, and supervised random walk was used for citation recommendation. 

However, all the existing methods cannot be used to address scientific formula evolution for two reasons. First, one complex formula (in a paper) may implicitly associate with different kinds of topics, and these topics may not appear in the formula context (text information is not complete). Second, scholarly publication citation information may not be sufficient for formula understanding, e.g., ``LDA'' studies do not necessarily cite ``Beta distribution'' or ``Bayesian inference'' foundations (that can be important to help readers to understand the formula). 

\textbf{Formula Search and Layout Mining:} Formula search is an important area in information retrieval. Recently, National institute of informatics Testbeds and Community for Information access Research (NTCIR) developed an evaluation collection for mathematical formula search with the aim of facilitating and encouraging research in formula search and its related fields \cite{aizawa2014ntcir}. For formula search, one of the main challenges should be formula information extraction, namely how to convert formulae into terms which were utilized to build the index. The same with text tokenizer, a text-based category of tokenizers was employed for formula retrieval and formula layout mining \cite{miller2003technical, miner2007approach, mivsutka2008extending}. Different from plain text, formulae were highly structured. They can be expressed and parsed as tree structures. Thus, tree-based methods were the most important tokenization approach in recently proposed formula search systems \cite{kohlhase2006search, schellenberg2012layout, sojka2011indexing, wang2015wikimirs}. However, formula understanding and formula evolution mining are novel problems, and existing formula search methods cannot be directly applied.
\vspace{-2ex}\section{Conclusion}

In this study, we propose a novel problem-Mathematics Content Understanding-to assist readers to better understand and consume the math-content in scientific publications by leveraging Formula Evolution Map (FEM) and high-quality OERs. By using the PRMA cyberreading system, students/scholars can easily highlight a target formula in a PDF reading, and the proposed algorithms can project the query formula to the formula(e) vertex(es) in FEM as well as recommend OERs to users. In the offline process, we extract formula evolution relations from a massive Wikipedia dump. Evaluation shows that formula relations on FEM are fundamental and enlightened for math-understanding. Most of the experiment participants find the proposed method/system can effectively help them better understand the math-content and readings in a cyberreading environment. Meanwhile, students reported that the formula highlight function was easy to use and the recommended OERs can be very useful for their understanding of the math-content. 

For the algorithm evaluation, we found that: first, the proposed OER recommendation via FEM mining is effective. It achieves the best performance for all evaluation metrics. Compared with the baselines without FEM assistance, the new model is significantly superior. Second, formula evaluation information is especially important for math-content understanding tasks, i.e., compared with the other comparison groups with partial FEM assistance, the proposed method with complete FEM information has an average increase of 8.2\% for P@3 and an average increase of 11.9\% for NDCG@3.

The methodological limitations of this work is that the parameter $\mathbf{\theta}$ of evolution probability calculation is treated equally without tuning. This is caused by two reasons. First, there is no existing formula evolution relations for training. Second, the number of scientific knowledge base documents and associated formulae is huge, and the time cost of direct parameter tuning can be very high. In the future, we will explore the combination of knowledge base and academic corpus in depth, and try to use more features (e.g., selected paper citation relation information) in the formula evolution mining. Meanwhile, we will propose more sophisticated optimization methods for FEM evolution probability parameter tuning. 
\vspace{-2ex}
\begin{acks}
The work is supported by the National Science Foundation of China (61472014), Guangdong Province Frontier and Key Technology Innovative Grant (2015B010110003, 2016B030307003) and the Opening Project of State Key Laboratory of Digital Publishing Technology.
\end{acks}
\bibliographystyle{ACM-Reference-Format}
\bibliography{refs} 

\end{document}